\pdfoutput=1
\documentclass[amsmath,superscriptaddress,aps,pra,twocolumn,floatfix]{revtex4-2}
\usepackage[T1]{fontenc}
\usepackage[utf8]{inputenc}
\usepackage{graphicx}
\usepackage{subfigure}
\usepackage{bm}
\usepackage{microtype}
\usepackage{amssymb,amsmath,amsbsy,amsgen,amsfonts}
\usepackage{multirow}
\usepackage{amsthm}
\usepackage{mathrsfs}
\usepackage{latexsym}
\usepackage{xcolor}
\usepackage{amstext}
\usepackage{txfonts}
\usepackage{pifont}
\usepackage{braket}
\usepackage{mathtools}
\usepackage[colorlinks=true,allcolors=blue]{hyperref}
\usepackage[title]{appendix}
\usepackage[british]{babel}
\usepackage{siunitx}
\usepackage{booktabs}

\allowdisplaybreaks[1]
\newcommand{\be}{\begin{equation}}
\newcommand{\ee}{\end{equation}}
\newcommand{\ba}{\begin{array}}
\newcommand{\ea}{\end{array}}
\newcommand{\bqa}{\begin{eqnarray}}
\newcommand{\eqa}{\end{eqnarray}}

\usepackage{threeparttable}
\newcommand{\cell}[2]{\parbox[t]{#1}{\raggedright #2}}

\begin{document}

\title{A divergent-beam surface plasmon resonance architecture for multiplexed malaria biosensing}

\author{A. S. Kiyumbi}
\email{amos.kiyumbi@udsm.ac.tz}
\affiliation{Department of Physics, Mathematics and Informatics, Dar es Salaam University College of Education, University of Dar es Salaam, P.O. Box 2329, Dar es Salaam, Tanzania}

\author{J. H. Hossea}
\affiliation{Department of Electronics and Telecommunication Engineering, Dar es Salaam Institute of Technology,P.O. Box 2958, Dar es Salaam, Tanzania}


\begin{abstract}
We present a numerical study of a divergent-beam Kretschmann surface plasmon resonance (SPR) platform for multiplexed malaria biosensing. A Powell-lens-generated angular fan enables camera-based angular interrogation of spatially separated regions of interest on a single Au film, thereby removing the need for mechanical scanning. The framework combines transfer-matrix modelling of the prism/Au multilayer with an effective-adlayer description of biomolecular binding at the biofunctional interface. As a representative dual-biomarker case, we consider {Plasmodium} lactate dehydrogenase (pLDH) and histidine-rich protein~2 (HRP-2). Benchmarking of the N-SF11/Au(45~nm) baseline against published water/glycerol data reproduces the characteristic resonance positions and yields a bulk angular sensitivity of \SI{73.2181}{\degree\per RIU}. With representative aptamer-like and antibody-like recognition layers, the relevant sensing states remain within \SIrange{54}{57}{\degree} and produce distinct, detector-resolvable responses. Combining the optical model with effective-medium and Langmuir binding descriptions gives model-based detection limits of approximately \SI{5.5}{ng.mL^{-1}} for HRP-2 and \SI{5.8e-2}{ng.mL^{-1}} for pLDH. These results support divergent-beam SPR as a viable architecture for quantitative multiplexed malaria biosensing.
\end{abstract}

\maketitle

\section{Introduction}
Surface plasmon resonance (SPR) remains one of the most mature label-free optical biosensing techniques because the plasmon resonance at a metal--dielectric interface is highly sensitive to refractive-index changes within the evanescent field. It therefore enables real-time measurements of adsorption, binding kinetics, and interfacial transformations without fluorescent or enzymatic labels~\cite{homola1999surface,homola2008surface}. For translational biosensing, however, raw sensitivity alone is rarely sufficient. Many clinically relevant problems are inherently multiplexed, require on-chip referencing, and benefit from internal comparison of distinct biomarkers measured under identical optical and fluidic conditions~\cite{rosa2022multiplexed}. In this context, SPR is especially attractive because a common optical platform can, in principle, support parallel biochemical capture, target/reference subtraction, and time-resolved analysis on a shared sensing surface~\cite{OBrien2001,Wang2019}.

This need is particularly clear in malaria diagnostics~\cite{hopkins2007comparison,yerlikaya2022dual}. Histidine-rich protein~2 (HRP-2) is often the more analytically sensitive and heat-stable malaria biomarker/antimalarial drug target, but it can persist after treatment and may fail in infections involving {pfhrp2/3} deletions. {Plasmodium} lactate dehydrogenase (pLDH), by contrast, is more closely associated with active infection because it tracks viable parasites and clears more rapidly after treatment, although pLDH assays are often less sensitive~\cite{Moody2002,Wongsrichanalai2007,li2017performance}. Measuring both biomarkers therefore offers a richer diagnostic picture than relying on either biomarker alone~\cite{jang2022comparison,lynch2022evaluation,yerlikaya2022dual}.

Several SPR formats have already moved towards this goal. SPR imaging and array-based platforms demonstrated early that many sensing elements could be monitored in parallel on a single chip~\cite{OBrien2001,Lee2006,Otsuki2010}, while later nanohole and nanoplasmonic architectures sought to simplify coupling optics and improve chip-level integration for multiplex assays~\cite{Lesuffleur2008,Im2012}. More recent work has further shown that simultaneous detection of two proteins on a common SPR surface is experimentally feasible~\cite{Lee2024}. The remaining challenge is not simply to place multiple receptors on one chip, but to do so in a format that preserves local biointerface sensitivity, tolerates bulk drift and non-specific adsorption, and remains compatible with compact, fast, and mechanically simple optical readout~\cite{Wang2019,Lee2024}.

A promising route is divergent-beam angular interrogation in the Kretschmann geometry. In this approach, a Powell lens produces a fan of incident angles, and a detector records the reflected angular spectrum in a single exposure, thereby avoiding mechanical prism-angle scanning while preserving angular interrogation of the SPR dip~\cite{Hossea2017,Netphrueksarat2022,Koresawa2022,Koresawa2023}. Because one detector axis can encode angle while the orthogonal axis resolves position across the chip, the architecture is naturally compatible with spatially separated regions of interest (ROIs). Although divergent-beam SPR has been developed for rapid angular interrogation, its use as a multichannel pathogen-sensing architecture remains largely unexplored. 
\begin{figure*}[t]
\centering
\includegraphics[width=0.95\textwidth]{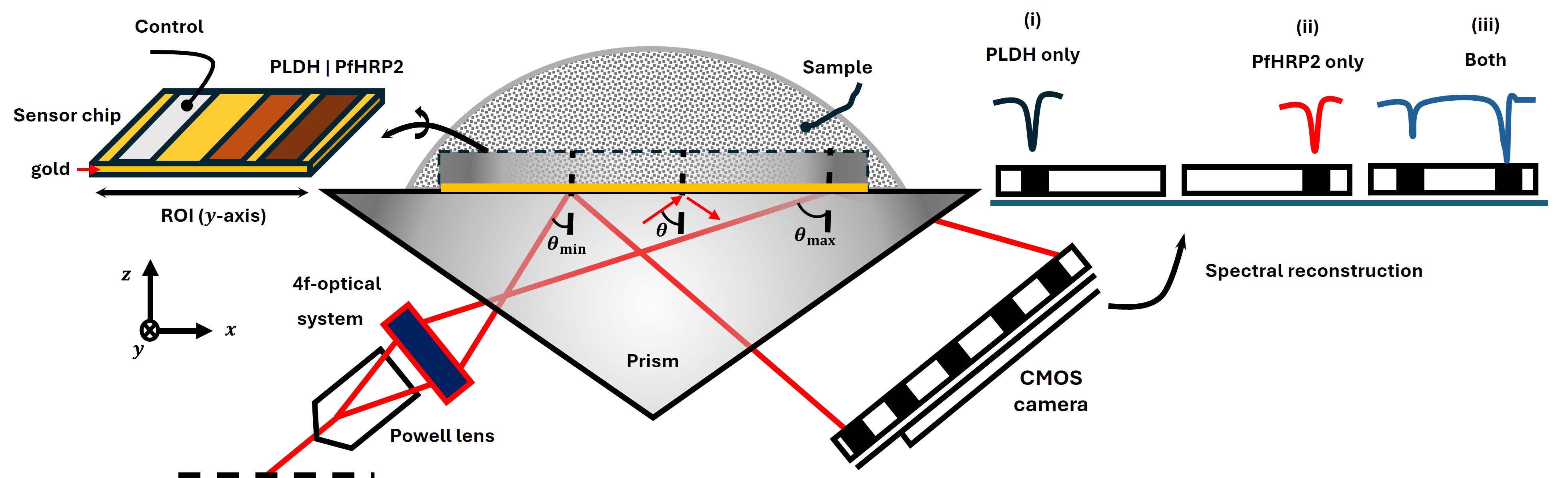}
\caption{Conceptual illustration of the divergent-beam Kretschmann SPR architecture used for multiplexed malaria biomarker sensing, adapted from Ref.~\cite{Netphrueksarat2022}. A Powell lens and a 4$f$ optical relay generate a controlled angular fan that interrogates a prism-coupled sensor chip without mechanical rotation. The Powell lens converts temporal angle scanning into spatial angle encoding. Spatially distinct regions of interest (ROIs) on the chip comprise a reference region and neighbouring capture regions selective to pLDH and PfHRP2. The reflected angular distribution from each ROI is acquired simultaneously on a linear or area detector (for example, a CMOS camera), enabling single-shot reconstruction of the SPR response in each sensing lane. The three example output states—(i) pLDH only, (ii) PfHRP2 only, and (iii) both biomarkers present—illustrate the basis for simultaneous discrimination and differential reference subtraction in the multiplexed assay.}
\label{fig:model_general}
\end{figure*}

Here we develop a numerical framework for multiplexed malaria biosensing in which pLDH and HRP-2 are detected simultaneously. The framework combines divergent-beam angular encoding, transfer-matrix modelling of the optical multilayer, and an effective-adlayer description that links receptor geometry, bound protein, and biomarker concentration to the final SPR observable. We use a simple N-SF11/Au(45~nm) structure adapted from Ref.~\cite{Netphrueksarat2022} as the baseline optical platform and show that, when combined with effective-adlayer modelling~\cite{Kiyumbi2025EffMicModel}, the divergent-beam geometry supports multiplexed acquisition and physically interpretable channel-resolved concentration responses.

The remainder of this manuscript is organised as follows. Section~II introduces the divergent-beam SPR concept and presents the theoretical framework used for angular encoding in the Kretschmann geometry. Section~III develops the multilayer optical model together with the effective-adlayer description for region-of-interest-resolved sensing. Section~IV presents the numerical results, including optical benchmarking, channel-resolved biomarker response, and model-based detection performance. Section~V discusses practical implementation considerations relevant to detector readout, microfluidic integration, and experimental deployment. Section~VI outlines the main limitations of the present study and possible future extensions. Finally, Section~VII summarises the main conclusions of this work.

\section{Theoretical framework}

\subsection{Angular encoding in the Kretschmann geometry}

In prism-coupled SPR, a $p$-polarised optical field incident through a high-index prism excites a surface plasmon (SP) at the metal--dielectric interface when the in-plane optical momentum $k_{\parallel}$ matches that of the plasmon mode $k_{\mathrm{SP}}$~\cite{raether2006surface,maier2007plasmonics}. For an incident angle $\theta$ in a prism of refractive index $n_p$, the conserved in-plane wavevector is
\begin{equation}
k_{\parallel}(\theta)=\frac{2\pi}{\lambda}n_p\sin\theta,
\end{equation}
with $\lambda$ the wavelength of light in free-space. In the simplest single-interface approximation, resonance occurs when $k_{\parallel}$ matches the real part of the surface-plasmon wavevector,
\begin{equation}
k_{\mathrm{SP}}=\frac{2\pi}{\lambda}\sqrt{\frac{\varepsilon_m\varepsilon_s}{\varepsilon_m+\varepsilon_s}},
\label{eq:ksp}
\end{equation}
where $\varepsilon_m$ and $\varepsilon_s$ are the metal and sensing-medium permittivities, respectively. In practice, the device is a multilayer system, so the resonance condition is more accurately obtained from the minimum of the full reflectance curve.

In the divergent-beam configuration considered here, a Powell lens generates a continuous range of incidence angles, $\theta\in[\theta_{\min},\theta_{\max}]$, as sketched in Fig.~\ref{fig:model_general}. The reflected intensity, which would in practice be imaged onto a detector, is calculated using the transfer-matrix method. Two operating modes are relevant. In resonance-tracking mode, the resonance angle $\theta_{\mathrm{res}}$ is extracted from the minimum of the angular reflectance profile. In fixed-angle mode, the reflected intensity is evaluated at one or more operating angles on the steep slope of the dip. For multiplexed sensing, resonance tracking is especially attractive because it provides a direct angular observable for each ROI without mechanical scanning~\cite{nguyen2015surface}.

\subsection{Divergent-beam interrogation in multiplexed SPR sensor}

Replacing mechanical prism-angle scanning with a divergent fan beam has several practical advantages. It reduces moving parts, shortens acquisition time, and makes the optical train more compatible with camera-based readout and multi-lane microfluidics~\cite{Hossea2017,Netphrueksarat2022}. The resulting precision of the extracted resonance does not depend only on pixel pitch; it is also governed by beam quality, optical aberrations, detector uniformity, digitisation depth, and the chosen resonance-extraction algorithm~\cite{Wang2019,Koresawa2023}. Nevertheless, divergent-beam SPR platforms have already been shown to deliver high angular resolution and strong refractive-index responsivity~\cite{Hossea2024BaTiO3,Hossea2024Design}.

This optical strategy is especially well suited to multiplexed assays. Rather than asking a single fixed-angle image to represent all biochemical channels simultaneously, the system retrieves an angle-resolved signature for each ROI under the same optical fan. Prior work on wide-angle and multichannel SPR imaging has established that spatially encoded angular readout is compatible with microfluidics, rapid acquisition, and advanced signal processing~\cite{Park2010AO,Liu2008AO,Zhang2014AO,Karabchevsky2011JNP,Watad2019OL,Vashistha2023Biosensors,Isaacs2015APL}. In the present study we use that foundation to model three neighbouring ROIs on a single Au film: a pLDH-selective region, an HRP-2-selective region, and a chemistry-matched reference region (refer Fig.~\ref{fig:model_general}).

\subsection{Transfer-matrix description of the optical platform}

Only $p$-polarised (or transverse magnetic---TM) light is considered, since conventional prism-coupled SPR is not excited by $s$-polarised light~\cite{MackayLakhtakia2022,raether2006surface,maier2007plasmonics}. For an $N$-medium stack, medium 1 is the prism, media 2 to $N-1$ are internal layers, and medium $N$ is the sensing medium. For an incident angle $\theta$ in the prism, as shown in Fig.~\ref{fig:model_general}, the normal wavevector component in layer $k$ is
\begin{equation}
k_{z,k}=k_0\left(n_k^2-n_1^2\sin^2\theta\right)^{1/2},
\qquad
k_0=\frac{2\pi}{\lambda},
\label{eq:kz}
\end{equation}
where $n_k$ is the complex refractive index of layer $k$. The phase thickness is $\beta_k=k_{z,k}d_k$, with $d_k$ the layer thickness, and the reduced optical admittance is
\begin{equation}
q_k^{(p)}=\frac{\left(n_k^2-n_1^2\sin^2\theta\right)^{1/2}}{n_k^2}
=\frac{\cos\theta_k}{n_k}.
\end{equation}
The characteristic matrix of layer $k$ is therefore
\begin{equation}
M_k=
\begin{pmatrix}
\cos\beta_k & -\,\dfrac{i}{q_k^{(p)}}\sin\beta_k \\[6pt]
-\,i\,q_k^{(p)}\sin\beta_k & \cos\beta_k
\end{pmatrix},
\label{eq:Mk}
\end{equation}
and the full transfer matrix is
\begin{equation}
M=\prod_{k=2}^{N-1}M_k=
\begin{pmatrix}
M_{11} & M_{12}\\
M_{21} & M_{22}
\end{pmatrix}.
\label{eq:Mglobal}
\end{equation}
The reflection coefficient of the multilayer then follows as
\begin{equation}
r_p(\theta)=
\frac{\left(M_{11}+M_{12}q_N^{(p)}\right)q_1^{(p)}-\left(M_{21}+M_{22}q_N^{(p)}\right)}
{\left(M_{11}+M_{12}q_N^{(p)}\right)q_1^{(p)}+\left(M_{21}+M_{22}q_N^{(p)}\right)},
\label{eq:rp}
\end{equation}
with reflectance
\begin{equation}
R_p(\theta)=|r_p(\theta)|^2.
\label{eq:Rp}
\end{equation}
For each chosen sensing state, the resonance angle $\theta_{\mathrm{res}}$ is obtained numerically from the minimum of $R_p(\theta)$ over the relevant angular window.
\section{Multilayer and effective-adlayer model}

\subsection{ROI stack definition and biomarker channels}

\begin{figure}[b]
\centering
\includegraphics[width=0.80\linewidth]{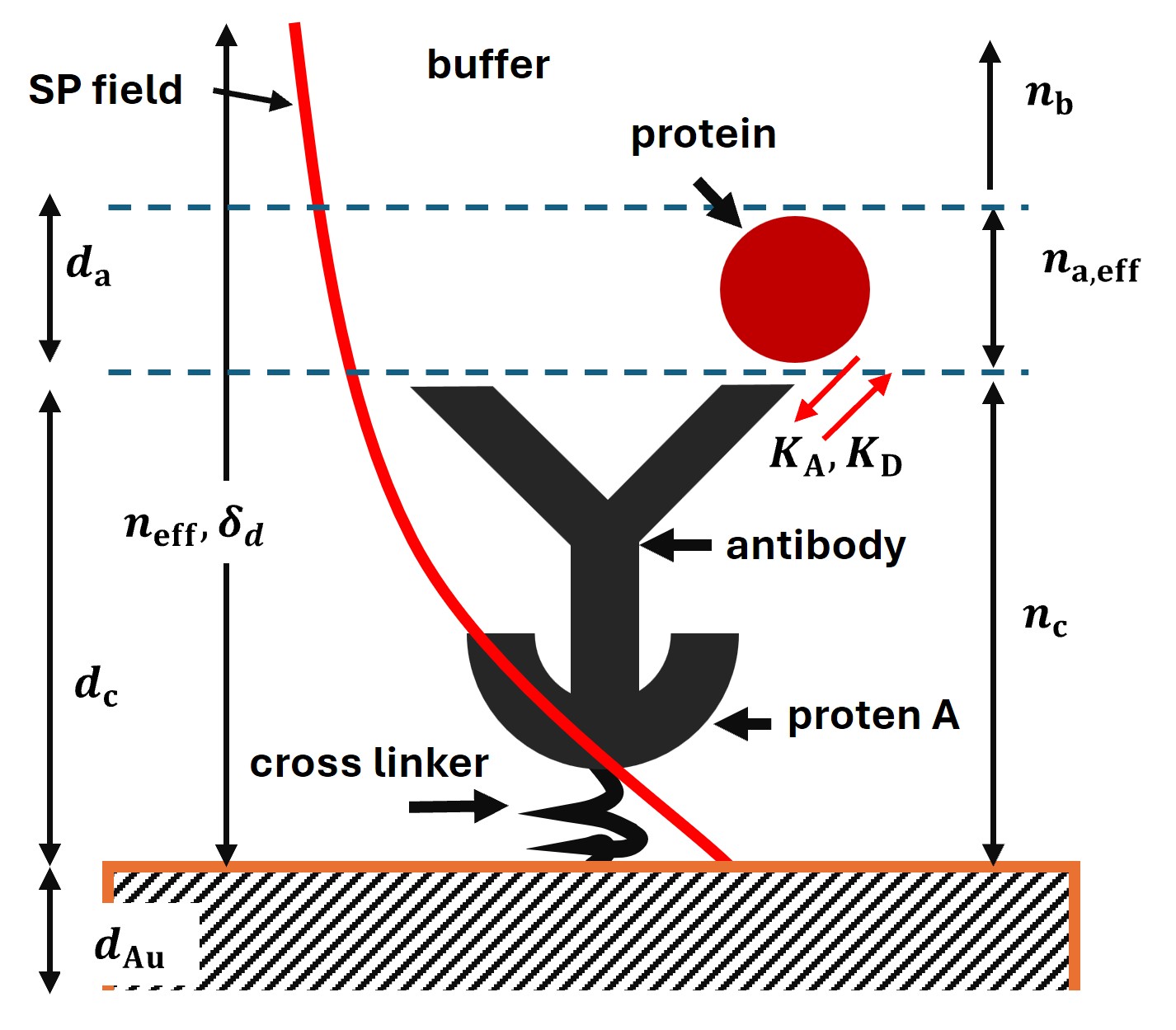}
\caption{Conceptual illustration of the divergent-beam Kretschmann SPR biointerface used for multiplexed malaria biomarker sensing. The evanescent surface-plasmon (SP) field decays exponentially from the Au surface into the buffer through the cross-linker, Protein~A, receptor layer, and bound analyte protein (red sphere). Key geometric and optical parameters are the receptor-layer thickness $d_c$, analyte-layer thickness $d_a$, gold-film thickness $d_{\rm Au}$, effective refractive indices $n_{\rm eff}$, $n_{a,{\rm eff}}$, $n_b$, and $n_c$, the plasmon penetration depth $\delta_d$, and the association/dissociation constants $K_A$ and $K_D$. This figure is not to scale.}
\label{fig:biointerface}
\end{figure}

Each ROI is modelled on the same optical base platform and differs only in the terminal biochemical layers. The multilayer sequence for ROI $i$ is
$
\text{prism}
\; \big|\;
\text{gold}\,(d_{\mathrm{Au}})
\; \big|\;
\text{bio-interface}\,(d_{\mathrm{c},i})
\; \big|\;
\text{protein adlayer}\,(d_{\mathrm{a},i})
\;~\big|~\\~\text{buffer (e.g., PBS)}.
$
As shown in Fig.~\ref{fig:biointerface}, $d_{\mathrm{Au}}$ is the Au film thickness, while $d_{c,i}$ denotes the fixed baseline chemistry present before target capture, including linker chemistry, the immobilisation scaffold (for example, Protein~A), and the receptor layer. Protein~A (from {Staphylococcus aureus}) is widely used as an intermediate layer for oriented IgG immobilisation on biosensor surfaces because it binds the Fc region of IgG with high affinity and helps to preserve Fab accessibility towards the solution phase~\cite{Wang2004,Gao2022,Choe2016}. The additional contribution due to target capture is represented by a thin analyte-containing adlayer of thickness $d_{a,i}$ and an effective refractive index $n^{(i)}_{\mathrm{a,\,eff}}$.

Representative values compatible with standard prism-coupled SPR chips are $d_{\mathrm{Au}}\approx\SIrange{45}{50}{nm}$ and a baseline biointerface thickness of order \SIrange{2}{20}{nm}, depending on the receptor chemistry~\cite{ghafoor2022optimization,fontana2006thickness}. In the present framework, the pLDH- and HRP-2-selective ROIs share the same optical base stack, while a third reference ROI is matched as closely as possible in surface chemistry but lacks specific capture. This arrangement is well suited to a divergent-beam platform because all ROIs experience the same angular fan and detector calibration, while the biomolecular state of each region can still be interpreted independently.

\begin{table*}[t]
\centering
\caption{Representative literature survey of malaria biomarker detection using aptamer- and antibody-based recognition. The table emphasises, for each target, the reported binders, transduction formats, and sample matrices most relevant to malaria biosensing.}
\label{tab:malaria_aptamer_antibody}
\footnotesize
\setlength{\tabcolsep}{3pt}
\renewcommand{\arraystretch}{1.12}
\begin{tabular}{lllll}
\toprule
\cell{2.8cm}{\textbf{Biomarker}} & \cell{2.2cm}{\textbf{Aptamer}} & \cell{3.0cm}{\textbf{Antibody}} & \cell{4.0cm}{\textbf{Detection platform}} & \cell{4.0cm}{\textbf{Sample type}} \\
\midrule
\cell{2.8cm}{{Plasmodium} lactate dehydrogenase}
& \cell{2.2cm}{pL1, pL2, 2008s~\cite{Lee2012pLDH,Cheung2018APTEC}; P38~\cite{Jain2016AuNP,Jain2016Graphene}; rLDH4, rLDH15, LDHp1, LDHp11~\cite{Frith2018Epitopic}}
& \cell{3.0cm}{anti-lactate dehydrogenase~\cite{Kiyumbi2026AlMetasurface}, anti-pan-lactate dehydrogenase~\cite{Piper1999ICpLDH}; 2CF5, 1G10~\cite{Linh2017PvLDH}; M1209063, M1709Pv2, M1709Pv1, M86550~\cite{Rogier2020PvLDH}; MBS498007, MBS498008~\cite{Mu2021ICPMS}}
& \cell{4.0cm}{electrochemical impedance spectroscopy~\cite{Lee2012pLDH,Lenyk2020Dual}; enzyme-linked oligonucleotide assay and aptamer-tethered enzyme capture~\cite{Cheung2018APTEC}; gold nanoparticle colourimetry and graphene-oxide electrochemistry~\cite{Jain2016AuNP,Jain2016Graphene}; rapid diagnostic test and sandwich enzyme-linked immunosorbent assay~\cite{Piper1999ICpLDH}; bead-based multiplex immunoassay~\cite{Rogier2020PvLDH}; surface-plasmon-polariton sensing on nanohole arrays~\cite{Lenyk2020Dual,Kiyumbi2026AlMetasurface}; plasmon-enhanced fluorescence on gold nanoparticle arrays~\cite{Minopoli2020NatComm,Minopoli2021RandomAuNP,Minopoli2022DoubleResonant}}
& \cell{4.0cm}{buffer~\cite{Lee2012pLDH,Jain2016Graphene}; serum~\cite{Lee2012pLDH}; whole blood and spiked whole blood~\cite{Minopoli2020NatComm,Minopoli2021RandomAuNP}; parasite culture and patient blood~\cite{Cheung2018APTEC,Linh2017PvLDH}; phosphate-buffered saline~\cite{Kiyumbi2026AlMetasurface}} \\
\hline
\cell{2.8cm}{{Plasmodium falciparum} histidine-rich protein~2}
& \cell{2.2cm}{B4~\cite{Chakma2018HRP2}; 2106s~\cite{Lo2021HRP2}}
& \cell{3.0cm}{anti-histidine-rich protein~2~\cite{Sharma2008JCM}; C1-13~\cite{Ravaoarisoa2010Recombinant}; D2, F9~\cite{Leow2014HRP2}; 1A5, 1C10~\cite{Kang2015HRP2RDT}; b10c1, Aa3c10~\cite{Verma2017Cterminal}; HCA159, HCA160~\cite{Sharma2008JCM}}
& \cell{4.0cm}{electrochemical impedance spectroscopy~\cite{Chakma2018HRP2}; square-wave voltammetry~\cite{Lo2021HRP2}; rapid diagnostic test~\cite{Kang2015HRP2RDT}; sandwich enzyme-linked immunosorbent assay~\cite{Kifude2008PfHRP2}; planar surface plasmon resonance~\cite{Sikarwar2014SPR}; fibre-optic surface plasmon resonance optrode~\cite{Loyez2021Optrode}}
& \cell{4.0cm}{serum and plasma~\cite{Kifude2008PfHRP2}; whole blood~\cite{Kang2015HRP2RDT}; culture supernatant / in vitro culture~\cite{Loyez2021Optrode}; phosphate-buffered saline spiked with purified target~\cite{Loyez2021Optrode}; clinical plasma~\cite{Parra1991PfHRP2}} \\
\bottomrule
\end{tabular}
\end{table*}

Table~\ref{tab:malaria_aptamer_antibody} highlights aptamers and antibodies as complementary receptor classes. Antibodies remain well established and often provide very strong affinity, whereas aptamers offer smaller physical dimensions, higher interfacial compactness, and convenient chemical modification~\cite{arshavsky2020aptamers,mpofu2025aptamers}. These distinctions matter optically because receptor thickness influences how efficiently the terminal perturbation is sampled by the plasmon evanescent field. We therefore use representative aptamer-like and antibody-like channels to study how receptor geometry and binding affinity co-determine multiplexed SPR performance.

\begin{table}[t]
\centering
\caption{Model biomarker--recognition pairs and reported affinities used to construct representative ROI-resolved concentration responses. Asterisks denote the two pairs we used for the detection-limit analysis.}
\label{tab:model_pairs}
\footnotesize
\begin{tabular}{llll}
\toprule
\cell{1.0cm}{\textbf{ROI}} & \cell{1.5cm}{\textbf{Biomarker}} & \cell{2.4cm}{\textbf{Recognition}} & \cell{2.4cm}{\textbf{Reported affinity}} \\
\midrule
ROI$_1$ & pLDH & \cell{2.4cm}{Aptamer pL1} & \cell{2.4cm}{\(K_D \approx 6.2\) nM~\cite{Cheung2018APTEC}} \\
ROI$_1$ & pLDH & \cell{2.4cm}{Aptamer 2008s} & \cell{2.4cm}{\(K_D \approx 43\) nM~\cite{Cheung2018APTEC}} \\
ROI$^{*}_1$ & pLDH & \cell{2.4cm}{Antibody 10C4D5} & \cell{2.4cm}{$K_D$ \(\sim 0.306\) nM~\cite{Kaushal2014PfLDH}} \\
\hline
ROI$_2$ & HRP-2 & \cell{2.4cm}{Antibody C1-13} & \cell{2.4cm}{\(K_D = 1.03\times10^{-10}\) M~\cite{Leow2014HRP2}} \\
ROI$_2$ & HRP-2 & \cell{2.4cm}{Antibody F9} & \cell{2.4cm}{\(K_D = 4.27\times10^{-11}\) M~\cite{Leow2014HRP2}} \\
ROI$^{*}_2$ & HRP-2 & \cell{2.4cm}{Aptamer 2106s} & \cell{2.4cm}{\(K_D \approx 29.53\) nM in PBS; \(22.59\) nM in serum~\cite{Lo2021HRP2,RoyeroBermeo2025Aptamers}} \\
\bottomrule
\end{tabular}
\end{table}

\subsection{Effective-index description of the biointerface}

To connect receptor geometry and target capture to the optical response, we use the effective-index formalism introduced previously in Ref.~\cite{Kiyumbi2025EffMicModel}. The baseline effective refractive index of ROI $i$ is written as
\begin{equation}
n_{\mathrm{eff}}^{(i)}=
n_{c,i}\!\left(1-e^{-2d_{c,i}/\delta_d}\right)
+n_b e^{-2d_{c,i}/\delta_d},
\label{eq:neff}
\end{equation}
where $d_{c,i}$ and $n_{c,i}$ are the thickness and refractive index of the receptor/biofunctional layer, $n_b$ is the buffer refractive index, and $\delta_d$ is the dielectric-side plasmon probe depth as shown in Fig.~\ref{fig:biointerface}. When molecules (e.g., protein) bind in ROI $i$ they form an analyte layer of thickness $d_{a,i}$. The induced change in effective refractive index is
\begin{equation}
\delta n_{\mathrm{eff}}^{(i)}=
\left(n_{a,\mathrm{eff}}^{(i)}-n_b\right)
\left(1-e^{-2d_{a,i}/\delta_d}\right)e^{-2d_{c,i}/\delta_d},
\label{eq:dneff}
\end{equation}
with $n_{a,\mathrm{eff}}^{(i)}$ the effective refractive index of the layer containing bound analyte. The corresponding optical response can be written as either an intensity change $\delta R_i=S_R\,\delta n_{\mathrm{eff}}^{(i)}$ or an angular shift
\begin{equation}
\delta\theta_{\mathrm{res},i}=S_\theta\,\delta n_{\mathrm{eff}}^{(i)},
\label{eq:dtheta_neff}
\end{equation}
where $S_\theta$ is the bulk angular sensitivity of the sensor. Equations~(\ref{eq:neff})--(\ref{eq:dtheta_neff}) expose an important distinction that is relevant in multiplexed plasmonic sensing. When the sensor response $\delta R$ is plotted against $n_{\mathrm{eff}}$, one probes the optical platform itself (non-functionalized) and therefore recovers the bulk sensitivity $S_\theta$ ---an intrinsic property of the sensor~\cite{homola1999surface}. When the response is plotted against $n_{a,\mathrm{eff}}$, the sensor sensitivity is modified by geometric and biochemical mapping factors~\cite{Kiyumbi2025EffMicModel}. The change of the effective index with respect to the analyte-layer refractive index is
\begin{equation}
\frac{\delta n_{\mathrm{eff}}}{\delta n_{a,\mathrm{eff}}}
=
\left(1-e^{-2d_a/\delta_d}\right)e^{-2d_c/\delta_d},
\label{eq:dneff_dnaeff}
\end{equation}
and therefore the local angular sensitivity $S^L_\theta$ with respect to the analyte-layer effective refractive index becomes
\begin{equation}
S^L_\theta = \frac{\delta\theta_{\mathrm{res}}}{\delta n_{a, \mathrm{eff}}}
=
S_{\theta}
\left(1-e^{-2d_a/\delta_d}\right)e^{-2d_c/\delta_d}.
\label{eq:local_sens}
\end{equation}
Eq.~\eqref{eq:local_sens} formalises how the terminal refractive-index perturbation is filtered by the finite penetration depth $\delta_d$ of the plasmon field. For two channels on the same optical platform with identical $d_a$ and $\delta_d$ but different receptor thicknesses, the ratio of local sensitivities is
\begin{equation}
\frac{S_{\theta,\mathrm{apt}}^{L}}{S_{\theta,\mathrm{ab}}^{L}}
=
\exp\!\left[
\frac{2(d_{c,\mathrm{ab}}-d_{c,\mathrm{apt}})}{\delta_d}
\right].
\label{eq:ratio_local_sens}
\end{equation}
Hence a thinner aptamer-like receptor layer composed of single-stranded nucleic acids~($< 5 \,\mathrm{nm}$), is expected to transmit the terminal perturbation more efficiently than a thicker antibody-like receptor layer~(roughly~$10-15 \, \mathrm{nm}$), even when both channels experience the same change in $n_{a,\mathrm{eff}}$~\cite{Kiyumbi2025EffMicModel,jung1998quantitative}.

\subsection{Effective microscopic model for concentration response}
To relate the terminal adlayer to biomarker concentration, the analyte layer is treated as a composite medium, i.e., a mixture of hydrated protein inclusions embedded in a host continuous medium---buffer. Using a Maxwell--Garnett description~\cite{MaxwellGarnett1904,Kiyumbi2025EffMicModel}, the effective refractive index of the layer $d_a$ in ROI $i$ is written as
\begin{equation}
\frac{\left(n_{a,\mathrm{eff}}^{(i)}\right)^2-n_{\mathrm{buffer}}^2}
{\left(n_{a,\mathrm{eff}}^{(i)}\right)^2+2n_{\mathrm{buffer}}^2}
=
V_i
\frac{n_{\mathrm{protein}}^2-n_{\mathrm{buffer}}^2}
{n_{\mathrm{protein}}^2+2n_{\mathrm{buffer}}^2},
\label{eq:MG}
\end{equation}
where $V_i$ is the volume fraction of bound protein and $n_{\mathrm{protein}}$ is the refractive index of the hydrated protein inclusion. The volume fraction is connected to bulk biomarker concentration $L_0$ through a Langmuir equilibrium model~\cite{Langmuir1918},
\begin{equation}
V_i(L_0^{(i)})=\frac{f_{\max,i}L_0^{(i)}}{K_{D,i}+L_0^{(i)}},
\label{eq:Langmuir}
\end{equation}
where $K_{D,i}$ is the dissociation constant of the chosen recognition pair (see Table~\ref{tab:model_pairs}) and $f_{\max,i}$ captures incomplete packing, hydration, and other departures from ideal close packing~\cite{Kiyumbi2025EffMicModel}. The angular response in ROI $i$ is then
\begin{equation}
\delta\theta_{\mathrm{res},i}(L_0^{(i)})=S_{\theta}\,\delta n_{\mathrm{eff}}^{(i)}(L_0^{(i)}),
\label{eq:dthetaL}
\end{equation}
and, in the multiplexed configuration, the working observable after common-mode subtraction is
\begin{equation}
\delta\theta^*_{\mathrm{res},i}(L_0^{(i)})=\delta\theta_{\mathrm{res},i}(L_0^{(i)})-\delta\theta_{\mathrm{res},\mathrm{ref},i}.
\label{eq:dthetaref}
\end{equation}
Finally, if $\sigma_\theta$ is the standard deviation of the blank angular response (here $\sigma_\theta$ is adopted from detector calibration as $\Delta\theta_{\text{pix}}$), a conventional $3\sigma$ concentration threshold is~\cite{Piliarik2009SPRlimits, Park2020SPR3sigma, Liu2022HSPRM}
\begin{equation}
\mathrm{LOD}^{(i)}
=\frac{3\sigma_\theta}
{\left|\delta\theta^*_{\mathrm{res},i}/\delta L_0^{(i)}\right|}=\frac{3\Delta\theta_{\text{pix}}}
{\left|\delta\theta^*_{\mathrm{res},i}/\delta L_0^{(i)}\right|}.
\label{eq:LOD}
\end{equation}
The LOD analysis used here is predictive: the optical platform, receptor geometry, and biomarker affinity are combined to estimate channel-specific performance under a pixel-calibrated angular noise floor. We take the angular noise floor to be $\sigma_\theta=\Delta\theta_{\mathrm{pix}}\approx1.1574\times10^{-3}\,^\circ/\text{pixel}$, following the reported detector calibration of the divergent-beam setup in Ref.~\cite{Netphrueksarat2022}.

\section{Results}
\begin{figure}[t]
\centering
\includegraphics[width=0.95\linewidth]{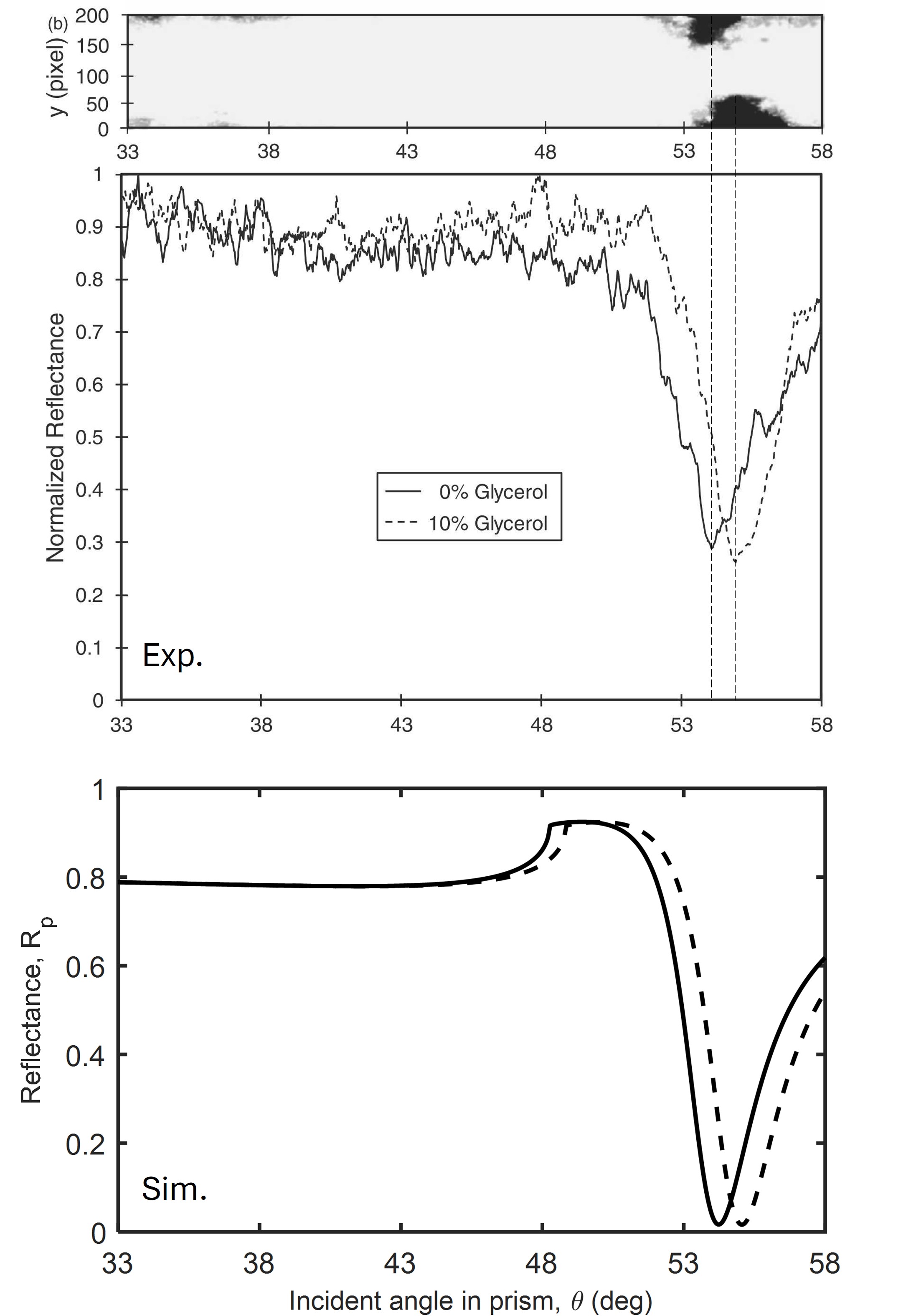}
\caption{Comparison of the experimentally measured and simulated SPR responses for water (0\% glycerol, refractive index 1.3317, solid line) and a 10\% glycerol solution (refractive index 1.3435, dashed line). The upper panel shows the recorded SPR reflectivity image as a function of vertical pixel position and incident angle, from which the one-dimensional traces in the middle panel were obtained by averaging along the vertical direction. The lower panel shows the corresponding transfer-matrix simulation. In both experiment and simulation, increasing the glycerol concentration shifts the SPR minimum to larger incident angles, consistent with the higher refractive index of the sensing medium. Experimental image and plots are extracted from Ref.~\cite{Netphrueksarat2022}.}
\label{fig:model_compr}
\end{figure}

Before introducing biofunctional layers, the baseline optical model was benchmarked against published N-SF11/Au(45~nm) divergent-beam measurements~\cite{Netphrueksarat2022} for water and 10\% glycerol, reproduced in Fig.~\ref{fig:model_compr}. The simulated minima agree well with experiment: for water ($n=1.3317$), the measured and simulated resonance angles are \SI{54.104}{\degree} and \SI{54.206}{\degree}, while for 10\% glycerol ($n=1.3435$) they are \SI{54.942}{\degree} and \SI{55.026}{\degree}, respectively. The discrepancies of only \SI{0.102}{\degree} and \SI{0.084}{\degree} indicate that the transfer-matrix model captures the dominant prism/Au/medium response with sufficient accuracy for design analysis. Equally importantly, the benchmark fixes a practical angular working region for the simple N-SF11/Au platform that is fully consistent with the demonstrated architecture~\cite{Netphrueksarat2022}.

\subsection{Bulk response of the baseline prism/Au stack}

\begin{figure}[t]
\centering
\includegraphics[width=0.95\linewidth]{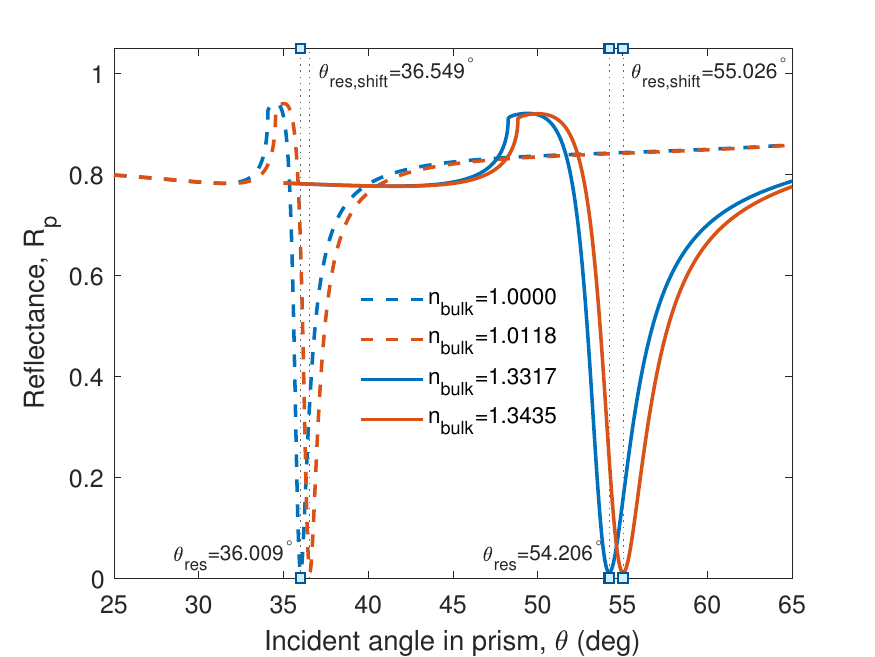}
\caption{Calculated angular reflectance curves for the non-functionalised N-SF11/Au(45~nm) structure for representative bulk media (air, low-index perturbation, water, and 10\% glycerol). The figure highlights the broad bulk-sensitive operating window of the simple prism/Au platform.}
\label{fig:bulk_response}
\end{figure}

\begin{figure}[t]
\centering
\includegraphics[width=0.95\linewidth]{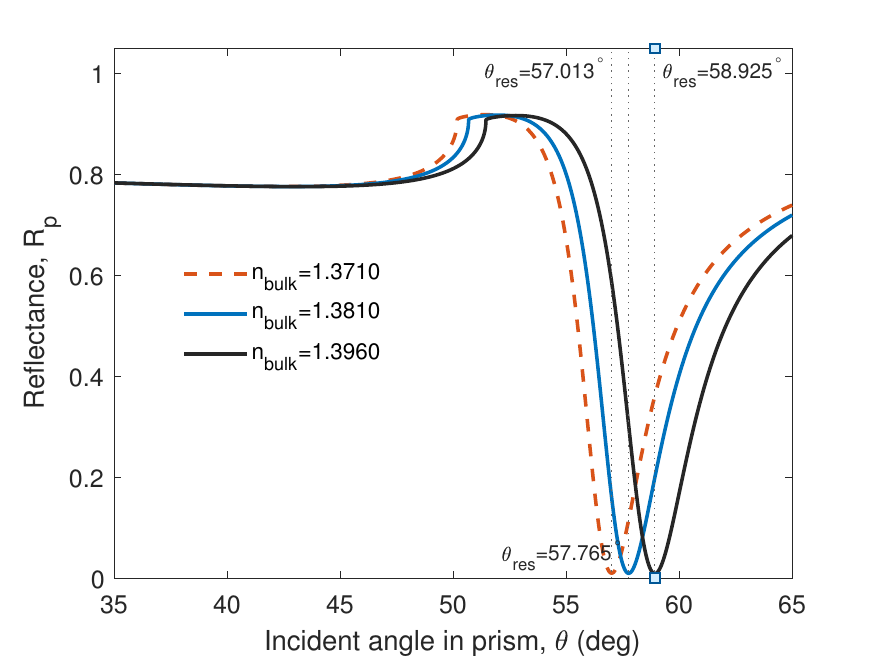}
\caption{Bulk sensing response for representative refractive indices associated with different {Plasmodium falciparum} intraerythrocytic stages. The simulated $p$-polarised reflectance curves for \(n_{\mathrm{bulk}}=1.3710\) (schizont), \(1.3810\) (trophozoite), and \(1.3960\) (ring) show a systematic shift of the resonance minimum towards higher incidence angles as the refractive index increases. The corresponding resonance angles are \(\theta_{\mathrm{res}}=\SI{57.013}{\degree}\), \(\SI{57.765}{\degree}\), and \(\SI{58.925}{\degree}\).}
\label{fig:malaria_response_stages}
\end{figure}

The bulk response of the baseline structure is shown in Fig.~\ref{fig:bulk_response} for representative media ranging from air to water and $10\%$ glycerol, while the malaria-specific bulk sensing is shown in Fig.~\ref{fig:malaria_response_stages}. We use reported refractive indices of healthy and {Plasmodium falciparum}-infected red blood cells~\cite{Park2008,Agnero2019} as a biologically motivated sensor calibration for the bulk-refractive-index regime. This is not intended as the main biosensing mechanism of our study; rather, it illustrates the part of index space over which the baseline prism/Au platform remains strongly responsive.

The refractive index of {P.\ falciparum}-infected red blood cells decreases progressively from the ring to the schizont stage, consistent with parasite maturation and haemoglobin consumption~\cite{Park2008,Agnero2019}. Using quantitative phase imaging, Park {et al.}~\cite{Park2008} reported mean cytoplasmic refractive indices of \(1.395 \pm 0.005\), \(1.383 \pm 0.005\), and \(1.373 \pm 0.006\) for the ring, trophozoite, and schizont malaria stages, respectively, compared with \(1.399 \pm 0.006\) for healthy red blood cells. This trend is supported by the single-cell measurements of Agnero {et al.}~\cite{Agnero2019}, who found corresponding values of approximately 1.396, 1.381, and 1.371 for the ring, trophozoite, and schizont malaria stages, respectively, as shown in Table~\ref{tab:rbc_refractive_index}.
\begin{table}[htbp]
\centering
\caption{Refractive~index~values for healthy and {P. falciparum}-infected red blood cells at different intraerythrocytic~stages.}
\label{tab:rbc_refractive_index}
\begin{tabular}{c c c c}
\toprule
\textbf{Refractive index range} & \textbf{Malaria stage} & \textbf{Index used} & \textbf{Ref.} \\
\midrule
$1.399 \pm 0.006$ & Healthy RBC  &  & \multirow{4}{*}{\cite{Park2008, Agnero2019}} \\
$1.395 \pm 0.005$ & Ring         & 1.3960 &  \\
$1.383 \pm 0.005$ & Trophozoite  & 1.3810 &  \\
$1.373 \pm 0.006$ & Schizont     & 1.3710 &  \\
\bottomrule
\end{tabular}
\end{table}

\begin{figure}[t]
\centering
\includegraphics[width=0.95\linewidth]{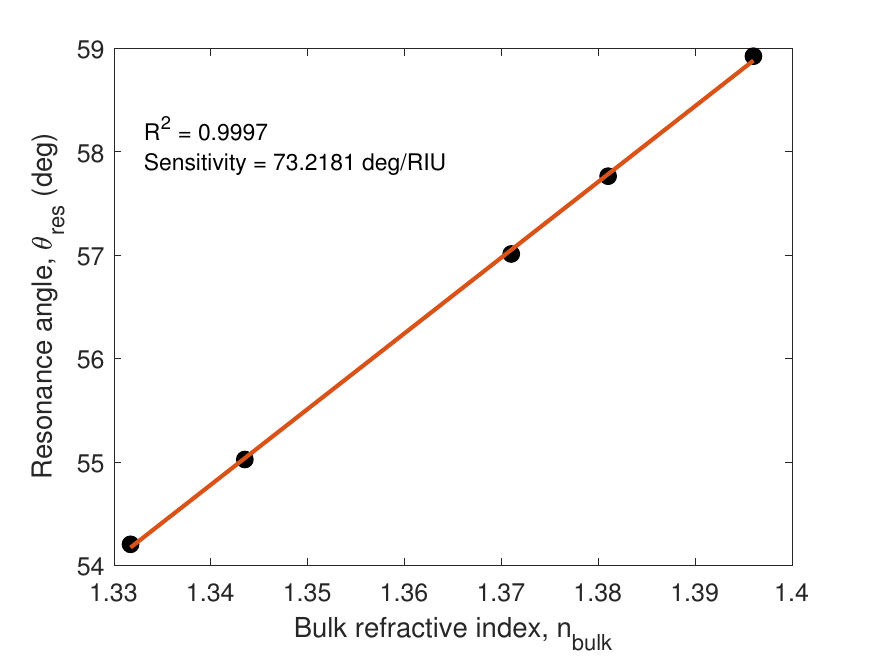}
\caption{Linear fit of resonance angle \(\theta_{\mathrm{res}}\) versus bulk refractive index \(n_{\mathrm{bulk}}\) for the representative malaria-stage refractive-index range. The fitted slope yields an angular bulk sensitivity of \(\SI{73.2181}{\degree\per RIU}\) with \(R^2=0.9997\).}
\label{fig:malaria_response_stages_sensitivity}
\end{figure}

\begin{figure}[t]
\centering
\includegraphics[width=0.95\linewidth]{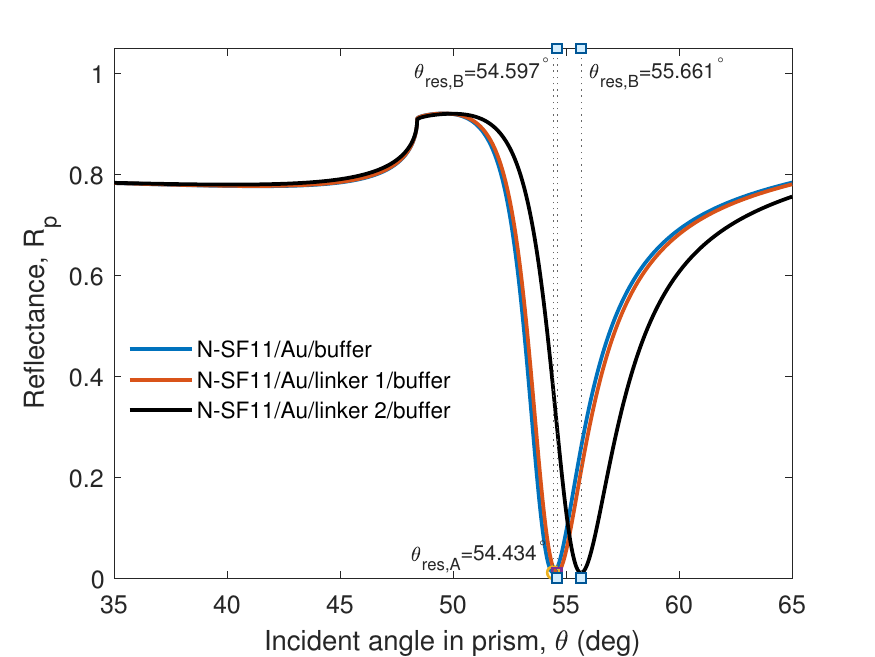}
\caption{Calculated effect of baseline biofunctionalisation on the resonance position for the N-SF11/Au platform. A thin aptamer-like biointerface (linker 1) and a thicker antibody-like biointerface (linker 2) are compared in buffer to show how the baseline operating point changes before analyte binding.}
\label{fig:linker_baseline}
\end{figure}
\begin{figure}[t]
\centering
\includegraphics[width=0.95\linewidth]{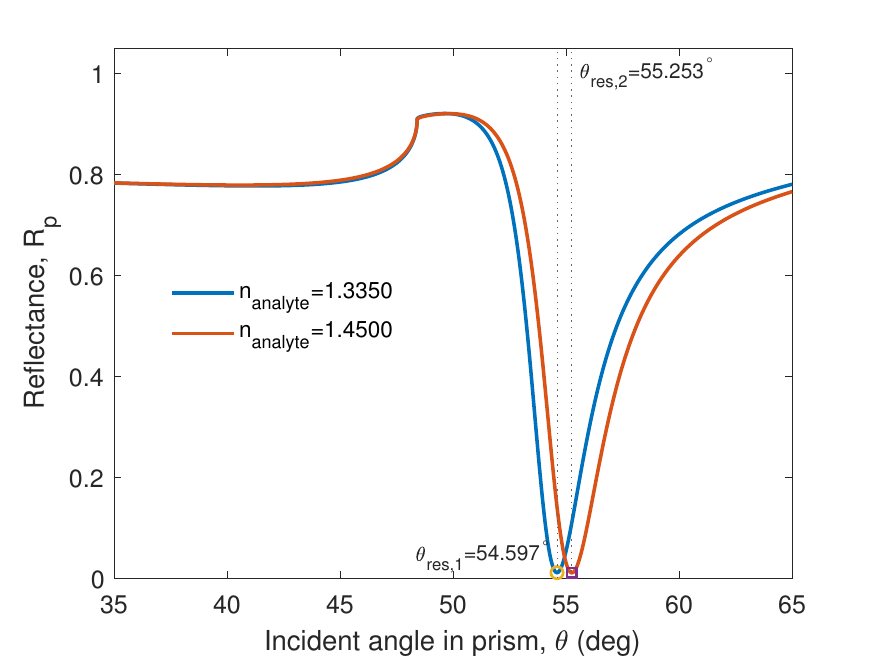}
\caption{Local-sensing response of the aptamer-like N-SF11/Au(45~nm) configuration before and after addition of the analyte layer. The analyte-induced angular shift is clearly resolved.}
\label{fig:local_response_apt}
\end{figure}

\begin{figure}[t]
\centering
\includegraphics[width=0.95\linewidth]{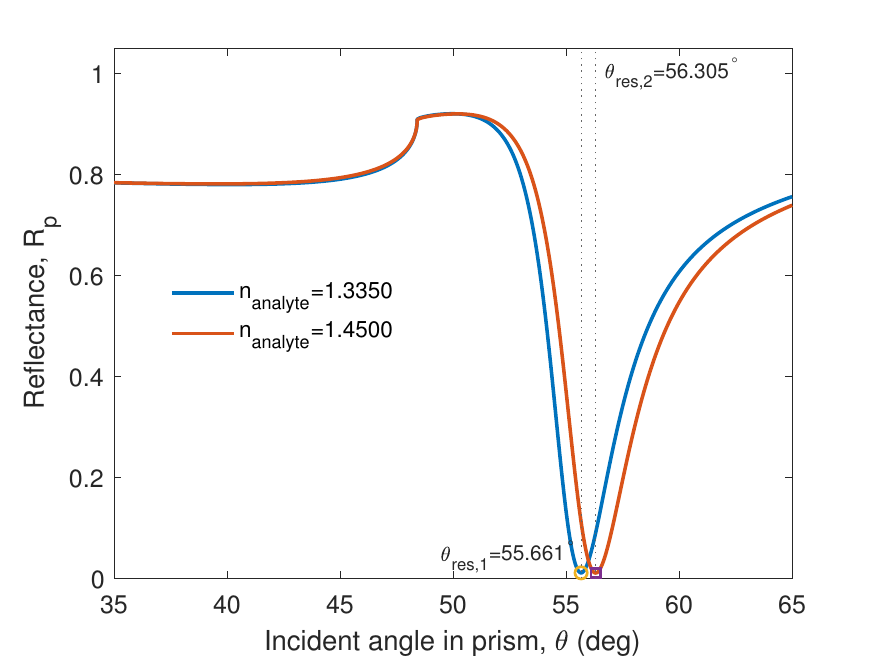}
\caption{Local-sensing response of the antibody-like N-SF11/Au(45~nm) configuration before and after addition of the analyte layer. The analyte-induced shift is again clearly resolved, with a final sensing state, $\theta_{\text{res}}= 56.305^\circ$, remaining within the operating window.}
\label{fig:local_response_ab}
\end{figure}
\begin{figure}[t]
\centering
\includegraphics[width=0.95\linewidth]{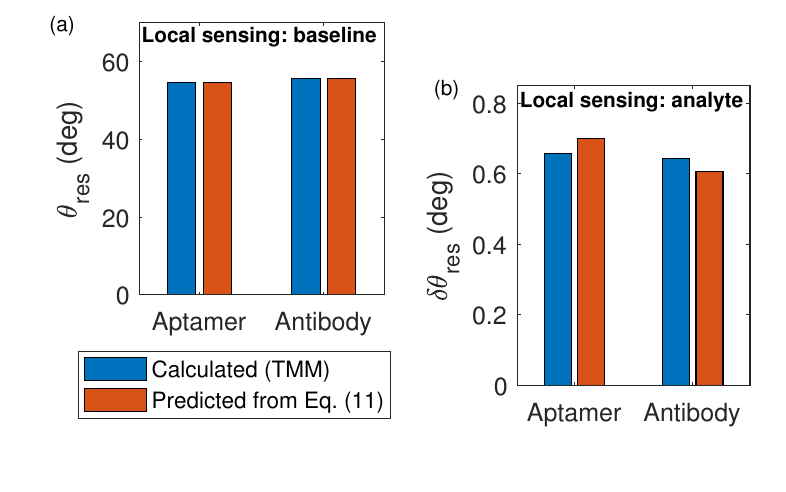}
\caption{Comparison between the resonance angles obtained from transfer-matrix simulations and those predicted by the reduced analytical model for the aptamer- and antibody-based sensing channels. Panel~(a) shows the baseline resonance angles after receptor functionalisation for \(n_{a,{\mathrm{eff}}}=1.335\) (no bound protein), while panel~(b) shows the corresponding sensor angular shifts \(\delta\theta_{\mathrm{res}}\) induced by analyte binding for \(n_{a,{\mathrm{eff}}}=1.45\).}
\label{fig:analytical_vs_tmm}
\end{figure}

From Fig.~\ref{fig:malaria_response_stages_sensitivity}, the bulk angular sensitivity of the baseline N-SF11/Au stack is \SI{73.2181}{\degree\per RIU}. This value is useful in two distinct senses. First, it confirms that the platform remains highly responsive to refractive-index changes in the angular range accessible to divergent-beam readout. Second, it provides the calibration bulk sensitivity that enters the effective-adlayer framework developed above. The subsequent local and concentration-dependent responses should therefore be read not as alternatives to bulk sensitivity, but as physically mapped versions of that same optical response once receptor geometry and biomolecular binding are included.

\subsection{Local biointerface response with receptor-like channels}
\begin{figure*}[t]
\centering
\includegraphics[width=\linewidth]{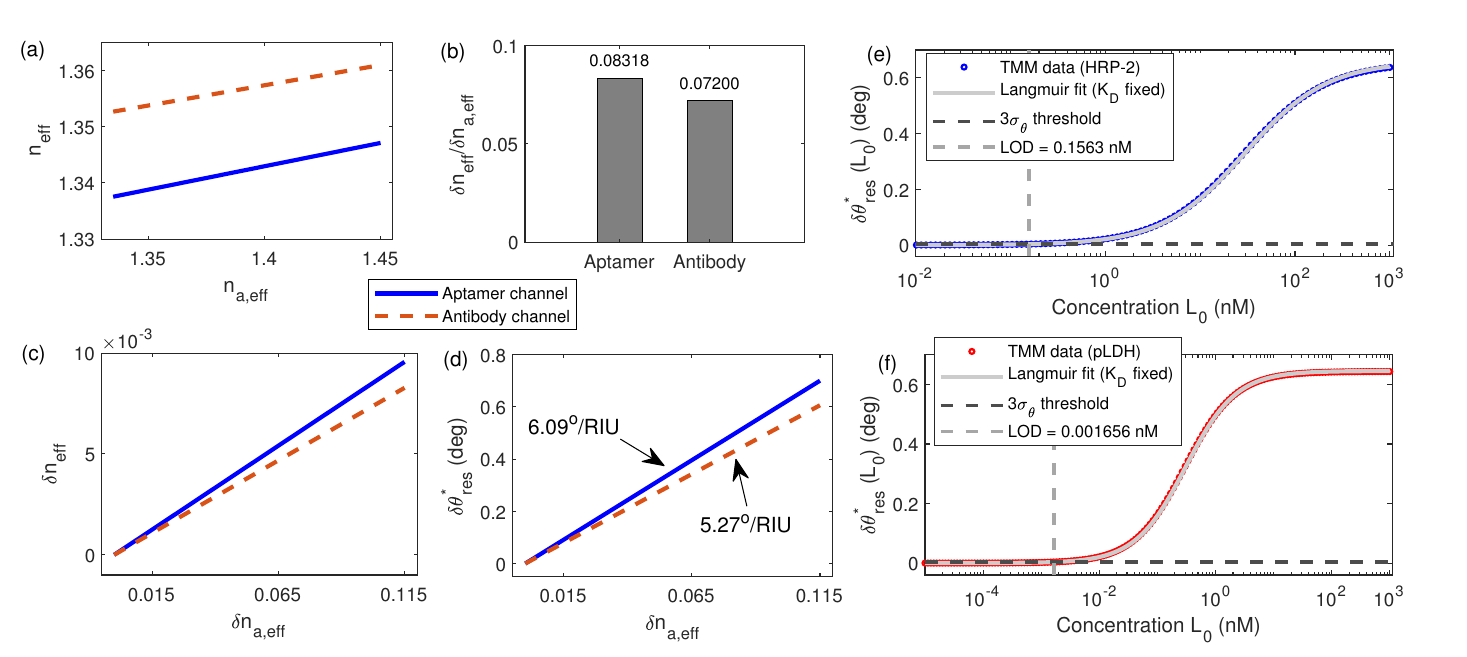}
\caption{Model-based comparison of local sensing in the aptamer- and antibody-functionalised SPR channels. (a) Effective interfacial refractive index, \(n_{\mathrm{eff}}\), versus the effective refractive index of the bound analyte layer, \(n_{a,\mathrm{eff}}\) based on Eq.~\eqref{eq:dneff_dnaeff}. (b) Near-field weighting factor, \(\delta n_{\mathrm{eff}}/\delta n_{a,\mathrm{eff}}\), for the two channels. (c) Induced effective-index change, \(\delta n_{\mathrm{eff}}\), as a function of \(\delta n_{a,\mathrm{eff}}\). (d) Corresponding angular response, \(\delta\theta^*_{\mathrm{res}}\) against \(\delta n_{\mathrm{eff}}\). Slopes in panel (d) correspond to the local angular sensitivity of the functionalised sensor platform. (e) Concentration-response curve for the HRP-2/2106s aptamer channel, showing the transfer-matrix data (from Table~\ref{tab:PfHRP2_2106s_representative}), the fixed-\(K_D\) Langmuir fit, the \(3\sigma_\theta\) detection threshold, and the extracted LOD. (f) Corresponding concentration-response curve for the pLDH/10C4D5 antibody channel (data from Table~\ref{tab:pfLDH_10C4D5_representative}).}
\label{fig:local_combined}
\end{figure*}

We next introduced representative biofunctional layers to determine how the same baseline optical stack behaves once receptor geometry and target capture are included. A related quantitative treatment was reported in Ref.~\cite{jung1998quantitative}. Here we consider two limiting receptor classes: a thin aptamer-like interface of \SI{2}{nm} and a thicker antibody-like interface of \SI{15}{nm}~\cite{arshavsky2020aptamers,mpofu2025aptamers}. Figure~\ref{fig:linker_baseline} shows that baseline biofunctionalisation shifts the resonance as expected while keeping the buffered operating points within a relatively compact angular interval. In buffer ($n=1.335$), the baseline states remain clustered near \SIrange{54}{56}{\degree}, which is favourable for simultaneous camera-domain acquisition of neighbouring ROIs.

After addition of a representative terminal analyte layer, the aptamer-like channel shifts from \SI{54.597}{\degree} to \SI{55.253}{\degree}, while the antibody-like channel shifts from \SI{55.661}{\degree} to \SI{56.305}{\degree}, as shown in Figs.~\ref{fig:local_response_apt} and \ref{fig:local_response_ab}. Both channels therefore produce clear analyte-induced shifts. The key physical point is that receptor thickness controls how strongly the terminal adlayer is sampled by the SP evanescent field. When the response is parameterised against \(n_{a,\mathrm{eff}}\), the corresponding local angular sensitivities, $\delta\theta_{\mathrm{res}}/\delta n_{a,\mathrm{eff}}$, are approximately \SI{6.09}{\degree\per RIU} for the aptamer-like channel and \SI{5.27}{\degree\per RIU} for the antibody-like channel. The thinner interface therefore transmits the local perturbation more efficiently, consistent with Eq.~\eqref{eq:ratio_local_sens}. These local sensitivities do not replace the bulk sensitivity of the platform; rather, they represent the bulk response after it has been filtered by the finite sampling of the bound-protein layer. The bulk angular sensitivity, which is an intrinsic property of the optical platform, is defined as
\begin{equation}
S_\theta = \delta\theta_{\text{res}}/\delta n_{\mathrm{eff}} = \left(\frac{\delta n_{\mathrm{eff}}}{\delta n_{a,\mathrm{eff}}}\right)^{-1}\times\frac{\delta\theta_{\text{res}}}{\delta n_{a,\mathrm{eff}}},
\label{local_linking}
\end{equation}
where ${\delta n_{\mathrm{eff}}}/{\delta n_{a,\mathrm{eff}}}$ describes how a change in the local analyte-layer refractive index maps onto the effective refractive index seen by the plasmonic transducer (see Fig.~\ref{fig:local_combined}(a,b)), while ${\delta\theta_{\mathrm{res}}}/{\delta n_{a,\mathrm{eff}}}$ quantifies the resulting local angular sensitivity. Evaluating the final two factors in Eq.~\eqref{local_linking} from Fig.~\ref{fig:local_combined}(b,d), namely $0.0831^{-1}\times 6.09$ for the aptamer channel and $0.07200^{-1}\times 5.27$ for the antibody channel, gives $S_\theta\approx 73^\circ/\text{RIU}$ for both cases. This is consistent with the independently calculated bulk value in Fig.~\ref{fig:malaria_response_stages_sensitivity}, $S_\theta=\SI{73.2181}{\degree\per RIU}$.

\begin{table}[t]
\centering
\caption{Representative local-sensing values for the PfHRP2/2106s aptamer channel used in the LOD analysis of Eq.~\eqref{eq:LOD}. The effective refractive index \(\mathbf{n_{a,\mathrm{eff}}}\) is calculated using Eq.~\eqref{eq:MG} over a concentration range of $0-1000$~nM. Binding kinetic $K_D=29.53$~nM and $f_{\max}=1$.}
\label{tab:PfHRP2_2106s_representative}
\scriptsize
\setlength{\tabcolsep}{4pt}
\renewcommand{\arraystretch}{1.12}
\begin{tabular}{cccc}
\toprule
\(\mathbf{n_{a,\mathrm{eff}}}\) & \(\boldsymbol{\theta_{\mathrm{res}}}\) (deg) & \(\mathbf{L_0}\) (nM) & \(\boldsymbol{\delta\theta^*_{\text{res}}(L_0)}\) (deg) \\
\midrule
1.335000 & 54.597406 & 0.000000   & 0.000000 \\
1.335015 & 54.597499 & 0.004028   & 0.000092 \\
1.335078 & 54.597877 & 0.020461   & 0.000470 \\
1.335354 & 54.599531 & 0.092552   & 0.002124 \\
1.336582 & 54.606901 & 0.418648   & 0.009495 \\
1.342611 & 54.642939 & 2.126822   & 0.045532 \\
1.362910 & 54.762791 & 9.620403   & 0.165384 \\
1.403037 & 54.993307 & 43.516651  & 0.395900 \\
1.436238 & 55.178092 & 221.074074 & 0.580686 \\
1.446644 & 55.234983 & 1000.000000 & 0.637576 \\
\bottomrule
\end{tabular}
\end{table}

\begin{table}[t]
\centering
\caption{Representative local-sensing values for the PfLDH/10C4D5 antibody channel used in the LOD analysis of Eq.~\eqref{eq:LOD}. The effective refractive index \(\mathbf{n_{a,\mathrm{eff}}}\) is calculated using Eq.~\eqref{eq:MG} over a concentration range of $0-1000$~nM. Binding kinetic $K_D=0.306$~nM and $f_{\max}=1$.}
\label{tab:pfLDH_10C4D5_representative}
\scriptsize
\setlength{\tabcolsep}{4pt}
\renewcommand{\arraystretch}{1.12}
\begin{tabular}{cccc}
\toprule
\(\mathbf{n_{a,\mathrm{eff}}}\) & \(\boldsymbol{\theta_{\mathrm{res}}}\) (deg) & \(\mathbf{L_0}\) (nM) & \(\boldsymbol{\delta\theta^*_{\text{res}}(L_0)}\) (deg) \\
\midrule
1.335000 & 55.661465 & 0.000000   & 0.000000 \\
1.335027 & 55.661622 & 0.000072   & 0.000157 \\
1.335213 & 55.662727 & 0.000577   & 0.001261 \\
1.336509 & 55.670401 & 0.004134   & 0.008936 \\
1.346113 & 55.726951 & 0.033267   & 0.065486 \\
1.388182 & 55.968116 & 0.267715   & 0.306651 \\
1.435477 & 56.227572 & 2.154435   & 0.566107 \\
1.447726 & 56.292933 & 15.441136  & 0.631468 \\
1.449712 & 56.303463 & 124.262367 & 0.641998 \\
1.449964 & 56.304797 & 1000.000000 & 0.643332 \\
\bottomrule
\end{tabular}
\end{table}

\subsection{Predicted concentration response and detection limits}
The distinction between bulk sensitivity, local adlayer sensitivity, and concentration response is summarised in Fig.~\ref{fig:local_combined}. Panels~(a)--(d) show that the optical platform retains a large bulk slope when the response is expressed against \(n_{\mathrm{eff}}\), but that the mapping to \(n_{a,\mathrm{eff}}\) is filtered by the finite penetration depth of the plasmon. The aptamer-like channel therefore exhibits a larger \(\delta n_{\mathrm{eff}}/\delta n_{a,\mathrm{eff}}\) than the antibody-like channel, as seen in panels~(a) and~(b). Increasing the bound-protein adlayer index \(n_{a,\mathrm{eff}}\) produces a larger change in the effective refractive index \(n_{\mathrm{eff}}\) for the aptamer-like biointerface than for the antibody-like one, and consequently a larger local optical response. This is the local SP field-weighting effect discussed above. Once kinetic binding and receptor affinity are included, the resulting concentration-response curves in Fig.~\ref{fig:local_combined}(e,f) allow the detection limit to be estimated using the conventional $3\sigma$ criterion.

Sensor response and LOD are determined jointly by optical field weighting and by the binding affinity of the recognition pair in each channel. A channel with a somewhat smaller \(\delta\theta_{\mathrm{res}}/\delta n_{a,\mathrm{eff}}\) can still deliver a favourable \(\delta\theta_{\mathrm{res}}/\delta L_0\) if its recognition pair is sufficiently strong. This distinction is central to the present analysis: the pLDH antibody channel ultimately achieves the lowest predicted concentration threshold despite its thicker receptor layer, because its biochemical affinity compensates for the weaker local plasmon field weighting~\cite{nguyen2015surface}.

In the present parameter set, the HRP-2/2106s channel benefits from stronger SP field weighting but has the weaker recognition pair, whereas the pLDH/10C4D5 channel experiences a modest geometric penalty but benefits from substantially tighter binding. The concentration threshold is therefore not governed by local SP field weighting alone. Using Eq.~\eqref{eq:LOD} and adopting an angular noise floor equal to the reported detector calibration of the divergent-beam setup, \(\Delta\theta_{\mathrm{pix}}\approx1.1574\times10^{-3}\,^\circ/\)pixel~\cite{Netphrueksarat2022}, the resulting thresholds are approximately \SI{0.1563}{nM} for the HRP-2/2106s channel and \SI{0.001656}{nM} for the pLDH/10C4D5 channel. Using representative molecular masses of \(\sim\)\SI{35}{kDa} for HRP-2 and \(\sim\)\SI{34.8}{kDa} for pLDH, these correspond to about \SI{5.5}{ng.mL^{-1}} and \SI{5.8e-2}{ng.mL^{-1}}, respectively. 

Both values lie in a practically relevant low-ng\,mL\(^{-1}\) regime for malaria antigen detection, but they should be interpreted as model-based LOD estimates. In practice, experimental LODs should be extracted from repeated blank angular traces acquired on the realised platform. Representative aptamer- and antibody-based malaria biosensors, with emphasis on pLDH/PfLDH and PfHRP2 detection, are summarised in Table~\ref{tab:prisma_malaria_biosensors}.

\subsection{Channel separability and cross-talk}
A route to discriminating pLDH and HRP-2 with SPR is to implement spatial multiplexing on a single continuous gold film, such that the optical system interrogates multiple sensing regions in parallel under identical illumination and thermal conditions. Stripe widths in the $0.5-2$~mm range with $100-500$~$\mu$m spacing are typically straightforward to align and provide sufficient area while maintaining spatial separation in the camera image~\cite{linman2010interface,lee2012real}. The essential requirement is that ROIs remain resolvable as distinct ROIs in the reflected-intensity image~\cite{peterson2009surface}.

Therefore, we examine whether multiplexing is compromised either by direct plasmonic cross-talk between neighbouring ROIs or by insufficient separation of channel states in the detector domain. The relevant surface-plasmon length scales for the experimental prism/Au configuration are short. Using \(n_{\mathrm{Au}}=0.1728+3.4218i\) at \SI{635}{nm} and \(n_s=1.3317\)~\cite{Netphrueksarat2022}, one obtains \(\varepsilon_{\mathrm{Au}}\approx-11.679+1.183i\) and \(\varepsilon_d\approx1.773\). Equation~(\ref{eq:ksp}) then gives a lateral propagation length of approximately
\begin{equation}
L_{\mathrm{SP}}=\frac{1}{2\,\mathrm{Im}(k_{\mathrm{SP}})}\approx \SI{3.9}{\micro m},
\end{equation}
and a dielectric-side penetration depth of order
\begin{equation}
\delta_d=\frac{1}{\mathrm{Re}\!\left(\sqrt{k_{\mathrm{SP}}^2-\varepsilon_d k_0^2}\right)}
\approx \SI{180}{nm}.
\end{equation}
These scales are many orders of magnitude smaller than the lateral dimensions and separations of the patterned ROIs, so direct near-field plasmonic coupling between regions is negligible~\cite{nelson2001surface}.

Then the more relevant issue is detector-domain separability. In the demonstrated divergent-beam implementation~\cite{Netphrueksarat2022}, the sensor spans \SIrange{53}{56}{\degree} over 2592 pixels, corresponding to an angular calibration of \(\Delta\theta_{\mathrm{pix}}\approx 1.1574\times10^{-3}\,^\circ/\)pixel and a physical pixel pitch of \(\Delta x_{\mathrm{pix}}\approx\SI{2.20}{\micro m/pixel}\). For the representative states used here, the aptamer channel shifts by \(\Delta\theta_{\mathrm{apt}}=\SI{0.656}{\degree}\), corresponding to roughly 567 pixels or about \SI{1.25}{mm} on the detector. Even the smallest nearest-neighbour spacing between channels,
$
\Delta\theta_{\mathrm{cross}}=\SI{55.661}{\degree}-\SI{55.253}{\degree}=\SI{0.408}{\degree},
$
still corresponds to about 352 pixels or approximately \SI{0.775}{mm}. The two channels are therefore detector-resolvable in principle. Multiplex is thus determined primarily by the interplay of differential referencing, channel-specific concentration response, and assay drift rather than by direct plasmonic cross-talk.
\begin{table*}[t]
\centering
\begin{threeparttable}
\caption{Representative aptamer- and antibody-based malaria biosensors, with emphasis on pLDH/PfLDH and PfHRP2 detection.}
\label{tab:prisma_malaria_biosensors}
\footnotesize
\setlength{\tabcolsep}{4pt}
\renewcommand{\arraystretch}{1.18}
\begin{tabular}{lllllll}
\toprule
\cell{1.8cm}{\textbf{Biomarker}} & \cell{2.7cm}{\textbf{Recognition}} & \cell{4.2cm}{\textbf{Sensor platform}} & \cell{1.8cm}{\textbf{LOD}} & \cell{1.4cm}{\textbf{Mode}} & \cell{3.1cm}{\textbf{Buffer / blood / serum}} & \cell{1.0cm}{\textbf{Reference}} \\
\midrule
\cell{1.8cm}{pLDH / PfLDH} & \cell{2.7cm}{SELEX-derived ssDNA aptamers} & \cell{4.2cm}{Electrochemical impedance spectroscopy (EIS) aptasensor} & \cell{1.8cm}{1 pM} & \cell{1.4cm}{Experiment} & \cell{3.1cm}{Purified protein; malaria-positive blood samples} & \cell{1.0cm}{\cite{Lee2012pLDH}} \\
\cell{1.8cm}{PfLDH} & \cell{2.7cm}{pL1 aptamer} & \cell{4.2cm}{AuNP/CTAB colorimetric aptasensor} & \cell{1.8cm}{2.94 pM (buffer); 13.54 pM (human serum)} & \cell{1.4cm}{Experiment} & \cell{3.1cm}{Buffer; human serum} & \cell{1.0cm}{\cite{Lee2014}} \\
\cell{1.8cm}{PfLDH} & \cell{2.7cm}{2008s aptamer} & \cell{4.2cm}{Dual-mode gold nanohole aptasensor: EIS + SPR} & \cell{1.8cm}{1.4 pM (EIS); 23.5 nM (SPR)} & \cell{1.4cm}{Experiment} & \cell{3.1cm}{Tris--HCl buffer} & \cell{1.0cm}{\cite{Lenyk2020Dual}} \\
\cell{1.8cm}{PfLDH} & \cell{2.7cm}{LDHp11 aptamer} & \cell{4.2cm}{Flexible multielectrode electrochemical aptasensor} & \cell{1.8cm}{1.80 fM} & \cell{1.4cm}{Experiment} & \cell{3.1cm}{Spiked blood; spiked whole blood; {P.\ falciparum} in vitro cultures} & \cell{1.0cm}{\cite{FigueroaMiranda2021}} \\
\cell{1.8cm}{PfHRP2} & \cell{2.7cm}{B4 aptamer} & \cell{4.2cm}{EIS aptasensor} & \cell{1.8cm}{$\sim$3.15 pM} & \cell{1.4cm}{Experiment} & \cell{3.1cm}{Buffer; diluted serum / mimicked real sample} & \cell{1.0cm}{\cite{Chakma2018HRP2}} \\
\cell{1.8cm}{PfHRP2} & \cell{2.7cm}{2106s aptamer} & \cell{4.2cm}{Electrochemical aptamer biosensor using square-wave voltammetry} & \cell{1.8cm}{3.73 nM} & \cell{1.4cm}{Experiment} & \cell{3.1cm}{Human serum} & \cell{1.0cm}{\cite{Lo2021HRP2}} \\
\cell{1.8cm}{PfLDH} & \cell{2.7cm}{Anti-P-LDH monoclonal antibody + 2008s aptamer} & \cell{4.2cm}{Plasmon-enhanced fluorescence plasmonic biosensor} & \cell{1.8cm}{18 fM (0.6 pg\,mL$^{-1}$)} & \cell{1.4cm}{Experiment} & \cell{3.1cm}{Whole human blood (1:100 diluted)} & \cell{1.0cm}{\cite{Minopoli2020NatComm}} \\
\cell{1.8cm}{pfLDH} & \cell{2.7cm}{Anti-pfLDH IgG antibody} & \cell{4.2cm}{Aluminium nanohole SPR/EOT metasurface} & \cell{1.8cm}{1.3 nM (45.6 ng\,mL$^{-1}$)} & \cell{1.4cm}{Experiment} & \cell{3.1cm}{PBS} & \cell{1.0cm}{\cite{Kiyumbi2026AlMetasurface}} \\
\hline
\cell{1.8cm}{HRP-2} & \cell{2.7cm}{2106s aptamer-like recognition layer} & \cell{4.2cm}{Divergent-beam Kretschmann SPR} & \cell{1.8cm}{0.1563 nM (5.5 ng\,mL$^{-1}$)} & \cell{1.4cm}{Simulation} & \cell{3.1cm}{Model-based; PBS} & \cell{1.0cm}{This study} \\
\cell{1.8cm}{pLDH} & \cell{2.7cm}{10C4D5 antibody-like recognition layer} & \cell{4.2cm}{Divergent-beam Kretschmann SPR} & \cell{1.8cm}{0.001656 nM (5.8 $\times$ 10$^{-2}$ ng\,mL$^{-1}$)} & \cell{1.4cm}{Simulation} & \cell{3.1cm}{Model-based; PBS} & \cell{1.0cm}{This study} \\
\bottomrule
\end{tabular}

\begin{tablenotes}[flushleft]
\footnotesize
\item LOD values are retained in the units reported by the original studies. Where both molar and mass-concentration forms were available or readily inferred in the cited work, both are shown.
\end{tablenotes}
\end{threeparttable}
\end{table*}

The analytical model developed above, namely Eqs.~\eqref{eq:neff} and \eqref{eq:dneff}, reproduces the transfer-matrix results well and provides a useful physical interpretation. Taking \(n_c=1.45\), \(n_b=1.335\), \(d_{c,\mathrm{apt}}=\SI{2}{nm}\), \(d_{c,\mathrm{ab}}=\SI{15}{nm}\), \(\delta_d=\SI{180}{nm}\), and \(\theta_{\mathrm{bare}}=\SI{54.434}{\degree}\), Eq.~\eqref{eq:neff} predicts baseline aptamer- and antibody-channel resonance positions of approximately \SI{54.619}{\degree} and \SI{55.727}{\degree}, in close agreement with the simulated values of \SI{54.597}{\degree} and \SI{55.661}{\degree}. The angular deviations are only \SI{0.022}{\degree} and \SI{0.066}{\degree}, respectively.

For local sensing with \(d_a=\SI{8}{nm}\), comparable to the geometric size of pLDH and HRP-2, Eq.~\eqref{eq:dneff} predicts angular shifts of about \SI{0.700}{\degree} for the aptamer-like channel and \SI{0.606}{\degree} for the antibody-like channel. These compare reasonably with the transfer-matrix shifts of \SI{0.656}{\degree} and \SI{0.644}{\degree}, while reproducing the expected trend that the thinner interface yields the larger local SP field weighting, as described by Eq.~\eqref{eq:ratio_local_sens}. The reduced model is therefore more than a numerical convenience: it isolates the role of receptor thickness and analyte-layer sampling in a transparent way.

\section{Practical implementation considerations}
A practical way to discriminate two biomarkers in the same sample is to implement spatial multiplexing on a single continuous SPR gold film, so that the optical system interrogates multiple sensing regions in parallel under identical illumination and thermal conditions~\cite{chin2023dual,hu2016design}. Among the available patterning approaches, microfluidic patterning is generally the most reproducible and experimentally forgiving. In this approach, a PDMS patterning device containing parallel microchannels is conformally sealed onto the gold surface, and each channel delivers a distinct capture solution~\cite{pla2010pdms,wheeler2004poly,lee2001spr}: anti-pLDH antibody (or aptamer) to Region~1, anti-HRP-2 antibody (or aptamer) to Region~2, and a reference formulation to Region~C. After incubation, the chip is rinsed, the patterning device is removed, and a separate assay flow cell is attached for measurements. This two-stage workflow minimises cross-contamination and allows the capture layers to be formed under controlled local conditions. Alternative approaches such as sequential masking or microspotting can be useful for rapid prototyping, but they are often more sensitive to alignment errors and edge artefacts, especially in biological samples.\\

In blood-derived samples, the dominant limitation is rarely plasmonic sensitivity itself; rather, it is the combination of nonspecific adsorption, matrix-dependent bulk refractive-index variations, and baseline drift~\cite{chirco2025tutorial,finocchiaro2026tackling}. For this reason, the capture chemistry should be designed from the outset to maximise antifouling performance while retaining adequate receptor density and activity~\cite{finocchiaro2026tackling,qi2022development}. A widely used strategy for gold is a mixed self-assembled monolayer (SAM) comprising a functional thiol diluted with a PEG-thiol. The PEG component suppresses nonspecific protein adsorption, while the functional fraction enables covalent coupling of antibodies or aptamers to Regions A and B~\cite{jans2008stability,frederix2003enhanced}. After coupling, unreacted groups should be quenched and the entire surface blocked using a compatible blocking agent.

An alternative route is a biotin--streptavidin scaffold, which provides modular functionalisation and simplifies optimisation of receptor densities~\cite{hanson2024combining,hamming2021streptavidin}. Regardless of the chosen chemistry, three controls are essential: orthogonality tests using pLDH-only and HRP-2-only injections; blank-matrix measurements to quantify residual nonspecific response after referencing; and repeatability tests over multiple injections and rinses. For whole blood, dilution and/or plasma separation may improve stability and reproducibility, and the same sample-processing route should be carried through calibration and validation. The inferred concentrations remain meaningful under the intended operating conditions~\cite{masson2017surface,tukur2024plasmonic}.\\

\begin{figure*}[t]
\centering
\includegraphics[width=0.99\textwidth]{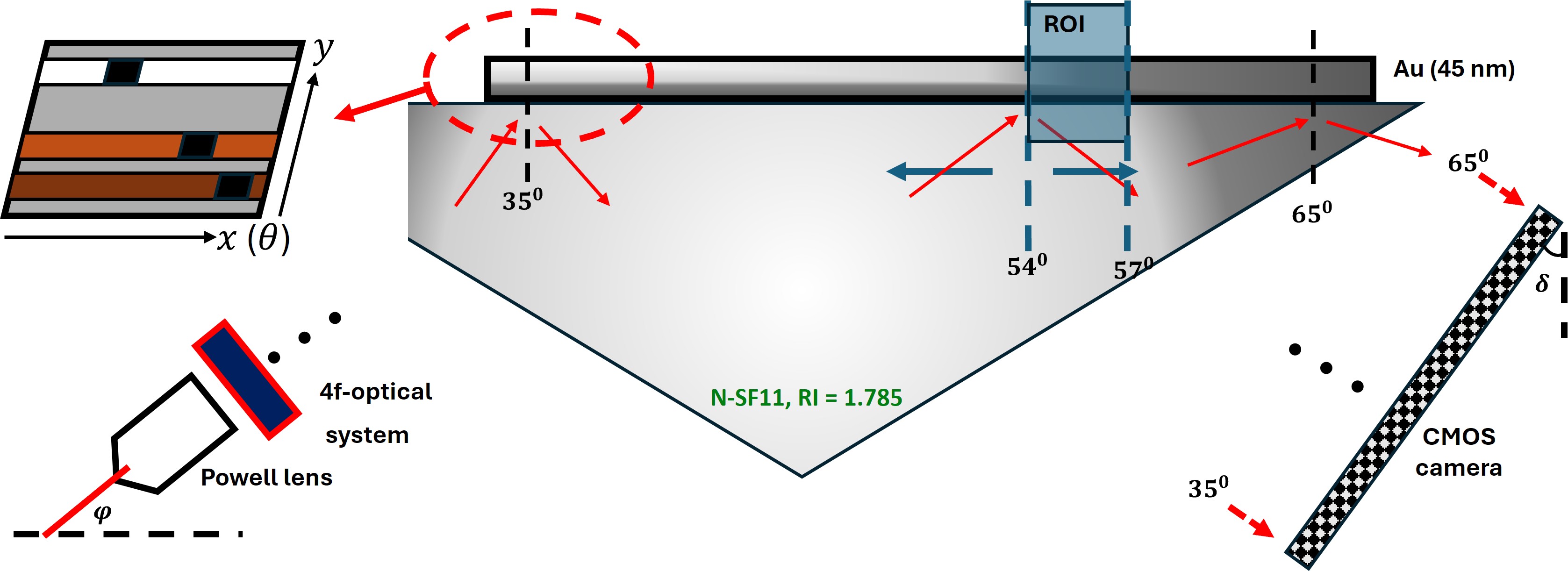}
\caption{Schematic of the practical divergent-beam SPR configuration used for multiplexed sensing. A $p$-polarised beam is expanded into an angular fan by a Powell lens and relayed by a 4$f$ optical system into an N-SF11 prism coated with a 45~nm Au film. Angular information is encoded along the horizontal detector axis ($x$), while spatial position on the SPR sensor chip is resolved along the vertical axis ($y$). The reflected SPR signal is collected by a tilted CMOS camera, allowing the response of selected ROIs to be extracted without mechanical angular scanning.}
\label{fig:practical_setup}
\end{figure*}

The numerical results suggest a practical and experimentally accessible implementation. The divergent-beam platform can be realised using a $p$-polarised He--Ne or diode light source near \SI{635}{nm}, a Powell lens with an approximately \SI{45}{\degree} fan angle, and a 4$f$ relay formed by two \SI{25}{mm} cylindrical lenses to image a compact angular fan into a \SI{15}{mm} N-SF11 equilateral prism coupled to a \SI{45}{nm} Au-coated glass chip~\cite{Netphrueksarat2022}. A useful starting geometry is a laser optical-axis inclination of \(\varphi=\SI{59}{\degree}\), producing an incidence span of roughly \SIrange{35}{58}{\degree} and an illuminated beam width of about \SI{1.5}{mm} on the Au surface. Figure~\ref{fig:practical_setup} summarises this layout.\\

For this application, an area-scan CMOS camera is preferable to a linear CCD because it captures the full two-dimensional angular--spatial distribution in a single exposure. A CMOS implementation is also typically more cost-effective than CCD-based alternatives, although CCD devices have historically offered advantages in noise and sensitivity~\cite{Zhang2014AO,bigas2006review}. In the demonstrated configuration of Ref.~\cite{Netphrueksarat2022}, camera tilt allowed the full sensor length to sample the \SIrange{53}{58}{\degree} interval with an angular calibration of \SI{1.1574}{mdeg/pixel}, producing experimental resonance dips near \SI{54}{\degree} and \SI{55}{\degree} for water and 10\% glycerol, respectively. These system-level details matter because they show that the simulated sensing states of the multiplexed biointerfaces remain within a realistically measurable angular window.

On the assay side, the reference ROI is central. For each sensing lane, the biomarker-specific observable can be written as
$
\delta R_{\mathrm{pLDH/HRP-2}}=R_{\mathrm{pLDH/HRP-2}}-R_C,
$
where \(R_\mathrm{pLDH}\), \(R_\mathrm{HRP-2}\), and \(R_C\) are the extracted observables from the pLDH, HRP-2, and reference ROIs, respectively. This subtraction is expected to suppress common-mode perturbations including bulk refractive-index drift, temperature fluctuations, and non-specific adsorption. The ROI layout should therefore be patterned so that the reference region is chemically matched as closely as possible to the active regions, differing only in the absence of specific capture.

\section{Limitations and outlook}

The present paper is a numerical study. The optical architecture has been benchmarked against published divergent-beam data, the multilayer response has been treated self-consistently, and the concentration response has been linked to explicit receptor affinities through an effective microscopic adlayer model. However, several elements still await experimental validation, including the proposed multiplexing scheme on a fabricated chip, the true angular noise floor, surface roughness, conformational heterogeneity, and lateral inhomogeneity of the receptor layer.\\ 

The next stage should therefore combine fabricated multi-ROI chips, chemistry-matched reference subtraction, repeated blank angular measurements, and controlled calibration in buffer and clinically relevant matrices. In parallel, the baseline N-SF11/Au structure used here can be extended to more strongly engineered multilayer stacks if additional near-field overlap with the terminal biointerface is required. Within that broader programme, the framework developed here remains useful because it separates optical response, receptor geometry, and biomolecular affinity into experimentally testable contributions.

\section{Conclusion}

We have presented a numerical study of a divergent-beam Kretschmann SPR architecture for multiplexed malaria biomarker sensing on a shared Au film. The platform combines spatially resolved ROIs, transfer-matrix modelling, and an effective-adlayer description that maps biomarker capture onto the optical response of the biofunctional interface. Benchmarking against published water/glycerol data confirms that the baseline N-SF11/Au(45~nm) platform is modelled accurately within the angular range relevant to existing divergent-beam instrumentation and yields a bulk sensitivity of \SI{73.2181}{\degree\per RIU}. When aptamer-like and antibody-like biointerfaces are introduced, both channels remain within the experimentally accessible angular window and produce detector-resolvable shifts.\\

The thinner aptamer-like interface transmits the local refractive-index perturbation more efficiently, whereas the thicker antibody-like interface can partially compensate for weaker SP field weighting through stronger biochemical affinity. As a result, both channels produce useful predicted concentration responses, and the model-based detection limits reach approximately \SI{5.5}{ng.mL^{-1}} for HRP-2 and \SI{5.8e-2}{ng.mL^{-1}} for pLDH under the adopted noise-floor. Together with the representative malaria biosensor literature summarised in Table~\ref{tab:prisma_malaria_biosensors}, these values place the divergent-beam SPR concept in a practically relevant low-concentration regime while retaining the key architectural advantage of simultaneous, detector-resolved multiplex interrogation on the same Au film. The present study therefore supports divergent-beam Kretschmann SPR as a credible route to multiplexed malaria biosensing and provides a grounded basis for the next experimental stage.

\bibliographystyle{apsrev4-2}
\bibliography{my_reference}

\end{document}